\documentclass[]{interact}

\usepackage{epstopdf}
\usepackage[caption=false]{subfig}
\usepackage{hyperref}
\usepackage{pdfpages}
\usepackage[numbers,sort&compress]{natbib}
\bibpunct[, ]{[}{]}{,}{n}{,}{,}

\theoremstyle{plain}

\theoremstyle{definition}

\theoremstyle{remark}

\begin{document}


\title{Leveraging LLMs for Persona-Based Visualization of Election Data}

\author{
\name{Swaroop Panda\thanks{Author Email: swaroop.panda@northumbria.ac.uk} and Arun Kumar Sekar}
\affil{Northumbria University}
}

\maketitle

\begin{abstract}
Visualizations are essential tools for disseminating information regarding elections and their outcomes, potentially influencing public perceptions. Personas, delineating distinctive segments within the populace, furnish a valuable framework for comprehending the nuanced perspectives, requisites, and behaviors of diverse voter demographics. In this work, we propose making visualizations tailored to these personas to make election information easier to understand and more relevant. Using data from UK parliamentary elections and new developments in Large Language Models (LLMs),  we create personas that encompass the diverse demographics, technological preferences, voting tendencies, and information consumption patterns observed among voters.Subsequently, we elucidate how these personas can inform the design of visualizations through specific design criteria. We then provide illustrative examples of visualization prototypes based on these criteria and evaluate these prototypes using these personas and LLMs. We finally propose some actionable insights based upon the framework and the different design artifacts.
\end{abstract}

\begin{keywords}
Election Data Visualization, User-Centered Visualization, User Personas, LLMs, Prototypes
\end{keywords}

\section{Introduction}
Electoral data and visualizations are prominently featured across various media publications \cite{nytimes2024,economist2024}. Political analysts examine with great scrutiny every poll, trend, and demographic shift in an effort to interpret the electorate's sentiment and forecast the electoral outcome. Media professionals construct narratives around these quantitative indicators, developing content that appeals to their readership and potentially influences public discourse \cite{alieva2023american}.
Furthermore, data scientists and visualization designers extensively engage in transforming raw statistical information into sophisticated graphics and interactive presentations. These visualizations range from choropleth maps that illuminate regional voting patterns to dynamic charts depicting campaign expenditure over time, thereby providing a comprehensive framework through which voters may engage with the electoral process \cite{silver2016fivethirtyeight}. Social media platforms serve as significant conduits for the dissemination of these visualizations, facilitating widespread discussion. Visual content including infographics and data-driven publications permeate digital spaces, influencing online discourse and shaping public perception regarding candidates and pertinent issues \cite{towner2017infographic}. In the contemporary digital environment, characterized by information abundance and diminished attention capacity, the significance of electoral data visualization in molding public opinion is substantial and cannot not be underestimated.

As visual representations wield significant influence over public opinion and have the capacity to sway decision-making processes \cite{lurie2007visual}, it becomes imperative to adopt a meticulous approach towards their design and dissemination within the realm of academia and beyond. The ethical imperatives surrounding visualization underscore the necessity for integrity and transparency throughout their lifecycle. Plaisant \cite{plaisant2004challenge} underscores the challenge of ensuring visualizations are not only technically sound but also ethically and informatively robust. This entails not merely presenting data, but rather curating it with rigorous fact-checking\cite{lo2022misinformed}, grounded research methodologies, and a steadfast commitment to portraying a comprehensive and unbiased perspective. Responsible data visualization \cite{correll2019ethical} thus pivots on empowering users to engage critically with information, enabling them to make informed decisions that are rooted in accuracy and devoid of bias. By prioritizing transparency, accuracy, and ethical considerations \cite{correll2019ethical} in the design and dissemination of visualizations, designers play a pivotal role in cultivating a populace that is well-informed and actively engaged in democratic processes.


Personas refer to fictional characters created to represent the different user types that might interact with a system, product, or service in a similar way\cite{mulder2006user}. These personas are meticulously crafted based on extensive research and data, ensuring they accurately reflect the demographics, behaviors, motivations, and goals of the target user group. They serve as proxies for real users, providing a tangible and relatable representation of otherwise abstract user data. The primary function of personas is to capture the peculiar characteristics of a group of people, thereby elucidating the commonalities within a group and highlighting the distinct differences between groups \cite{long2009real,miaskiewicz2011personas,mulder2006user,nielsen2021understanding}. This differentiation is crucial as it allows designers to tailor their solutions to meet the specific needs and preferences of different user segments, thereby enhancing the overall user experience. Moreover, personas play a critical role in mitigating bias and assumptions in the design process. By grounding design decisions in real user data, personas help ensure that the solutions developed are not based on the designers' subjective perspectives or preconceived notions, but rather on empirical evidence and genuine user needs \cite{pruitt2003personas}.

Visualizations can be catered to such personas. A benefit of catering to personas is that visualizations can be more user centered \cite{kay2016ish}; catering to user groups without stereotyping a large group of people.  Designers can then ensure that the information presented is relevant and meaningful to the intended user group, thereby enhancing user engagement and comprehension \cite{bottinger2020reflections,de2018does,lee2020reaching}. Furthermore, the persona-based approach facilitates ethical considerations in data visualization. By acknowledging the diverse perspectives and informational needs of different user segments, designers can mitigate potential biases in data presentation\cite{kim2021accessible}.  



In this work, we design and develop visualization from user personas. We use large language models (LLMs) to build the personas. We then derive design considerations for visualization from these personas, From these design considerations we develop the visualization prototypes. We then evaluate the visualization prototypes using LLMs. We use UK election datasets from the purpose of visualization and creating user personas.

The contributions of this work is basically fourfold;

\begin{enumerate}
 \itemsep0em 
    \item Three User Personas built using LLMs aimed at facilitating the visualization of election data, thereby fostering the development of user-centric visualizations.
    \item Design Considerations derived from these user personas and from visualization practice. 
    \item Design, Development \& Evaluation of the visualization prototypes for election data based on the Design Considerations.
    \item Actionable insights for visualization researchers derived from reflections on the methodological framework, design considerations and visualization prototypes. 
\end{enumerate}

\section{Background}
\subsection{Visualization in Elections}
Election outcomes evoke considerable public interest, compelling media entities \cite{graphics,silver2016fivethirtyeight,graphicdetail|theeconomist} to engage specialized personnel, including graphic designers and data analysts, in the creation of visual representations and comprehensive evaluations of electoral data. Infographics disseminated via social media channels yields discernible effects on the evaluation of presidential candidates. Visualization have been used to \textit{frame} US election results \cite{alieva2023american}. Visualization also helps developing competing narratives about election results \cite{solop2016data}. 

These infographics associated with political campaigns swiftly seize the audience's attention through aesthetically appealing graphical elements and typography, efficaciously presenting complex datasets in a comprehensible format, thereby facilitating expedited dissemination of information and amplifying message proliferation within condensed temporal intervals  \cite{towner2017infographic}. Visualizations wield substantial influence within the electoral sphere, as evidenced by their capacity to evoke heightened emotional responses, bolster trust levels, and potentially sway individuals' inclination to engage in electoral processes \cite{yang2023swaying}.

Various methodologies have been employed to visualize electoral data, encompassing a spectrum of visualization techniques. Comparative visualization approaches have been leveraged to juxtapose, superimpose voter preferences across local and central government elections \cite{panda2021comparing}. The design of dashboards integrating spatio-temporal and narrative visualization methods has emerged as a prominent strategy for elucidating the intricacies of US electoral results\cite{rao2021multi}. The application of sentiment analysis to location-based social media data pertaining to elections in the US and UK has facilitated the creation of compelling  visual representations \cite{yaqub2020location}.




In our study, we incorporate some methodologies from the big repository of visualization techniques utilized for electoral data analysis, including the integration of narrative visualizations. Concurrently, we remain cognizant of the consequential role played by such visualizations and infographics in potentially shaping individuals' voting comportment and political attitudes.
\subsection{Personas for Visualization}
User personas, a fundamental concept in user-centered design and marketing, are archetypical representations of target user groups, encapsulating their characteristics, needs, behaviors, and goals. These fictional characters serve as proxies for real users, aiding in the design and development of products, services, and marketing strategies tailored to specific user segments \cite{blomquist2002personas,miaskiewicz2011personas,pruitt2010persona}. Personas are important because they provide a human-centered framework for understanding and empathizing with users, guiding the design and development of products, services, or experiences. By creating fictional characters that represent distinct user archetypes, personas help teams visualize and internalize the needs, goals, behaviors, and preferences of their target audience\cite{miaskiewicz2011personas,long2009real}. These personas can be created by using qualitative, quantitative and mixed-methods \cite{salminen2020literature,mulder2006user,pruitt2003personas,jansen2022create}. 

User centered approaches have been used in visualization \cite{koh2011developing,goodwin2013creative,wassink2009applying}. They have been used for landmark-based navigation which is a natural concept for humans to navigate themselves through their environment \cite{elias2008user}. User centered approaches have also been used for visualizing uncertainity in predictive systems\cite{kay2016ish}. 









In this study, our aim is to adopt a user-centric methodology in visualizing electoral data, employing personas as pivotal drivers to inform the user centered design process. This approach underscores a nuanced understanding of user needs and preferences, facilitating the creation of tailored visualizations that resonate with diverse audience segments and enhance engagement with electoral information.

\subsection{LLMs in Visualization and Personas}


LLMs have been used in visualization research for a variety of purposes \cite{vazquez2024llms,lo2024good,hong2025llms}. They have been used to generate charts from natural language \cite{tian2024chartgpt}, to explain visualization recommendation \cite{wang2023llm4vis}, giving advice on visualization design \cite{kim2023good}, for visualization literacy \cite{hong2025llms} among others. As the capabilities of LLMs continue to grow, their potential applications in visualization research are expected to expand.

User Personas have also been created by LLMs \cite{de2023improved,huang2024unlocking}. By developing strategic prompts and guidelines, the researchers generated personas and compared them with those crafted by human experts. The findings revealed that LLM-generated personas were almost indistinguishable from human-written ones in terms of quality and acceptance, suggesting their potential to enhance efficiency in UX research and inspire stakeholder discussions \cite{schuller2024generating}. Furthermore, they afford a more profound, data-driven comprehension of distinct user segments, thereby enriching the analytical landscape pertaining to user-centered design methodologies \cite{huang2024unlocking}.

In this work we use LLMs to develop personas that can guide the visualization of election data. The objective does not pivot on the endeavor to replace human subjects with AI \cite{schmidt2024simulating,gerosa2024can}. Instead, the focal aim resides in leveraging expansive datasets, such as those encapsulated within LLMs, to build personas, which function as fictitious prototypes in their own right \cite{nielsen2021understanding}. This approach acknowledges the capacity of AI-driven personas to encapsulate and represent multifaceted dimensions of human behavior and cognition, thereby enriching analytical frameworks and augmenting comprehension within diverse domains.

\subsection{Politics in the UK}
In this study we use datasets from UK elections to visualize. The UK's political system has historically been dominated by two major parties: the Conservative Party (Tories) and the Labour Party. However, several smaller parties also maintain parliamentary representation and influence. The July 4, 2024 general election marked a significant political shift. Labour, won decisively, ending 14 years of Conservative rule and securing a strong parliamentary majority. The Conservative Party suffered historic losses with substantially reduced representation. Other notable outcomes included the Liberal Democrats achieving their best-ever result, the Green Party making gains amid growing environmental concerns, and Reform UK entering Parliament for the first time. The Scottish National Party experienced significant losses, while Sinn Féin became Northern Ireland's largest party. This election highlighted the UK's increasingly fragmented political landscape, with multiple parties gaining substantial support and challenging the traditional two-party system, prompting political strategy reassessments across the spectrum\cite{UKParliament2023}.

\section{Methodology}
\begin{figure}[!htb]
    \centering
    \includegraphics[width=0.55\textwidth]{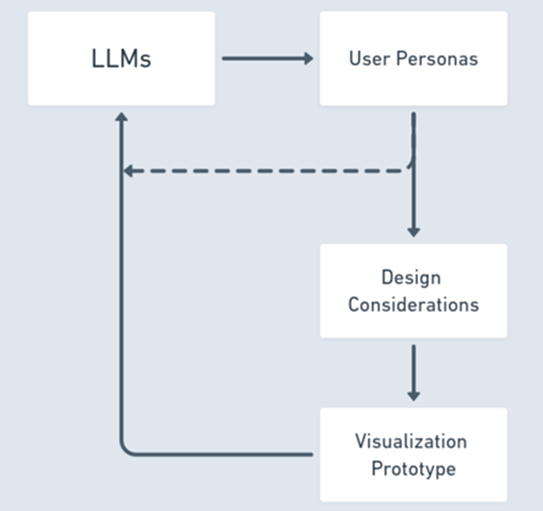}
    \caption{The Methodology}
    \label{fig:framework}
\end{figure}


The methodology outlined in this paper employs a systematic approach to designing and evaluating visualization prototypes using User Personas and LLMs. Initially, the process involves leveraging LLMs to generate user personas, which serve as foundational representations of target users. These personas are crafted to encapsulate diverse user characteristics, preferences, and behaviors, providing a robust basis for subsequent visualization development. Drawing from both the synthesized user personas and existing visualization research and practice, a set of design considerations is derived. These considerations integrate insights from existing literature with persona-specific requirements, ensuring that the resulting designs are both theoretically grounded and user-centered. Utilizing these carefully formulated considerations, visualization prototypes are then developed, embodying the principles and specifications identified in the prior stage. The final phase of the methodology involves an evaluation of these prototypes, again employing LLMs to assess the effectiveness, usability, and overall impact of the visualizations. This cyclical use of LLMs—spanning persona creation, prototype evaluation, and iterative refinement—represents an innovative application of AI in visualization research. By anchoring the design process in user personas and grounding it in scholarly and practical insights, this approach seeks to bridge the gap between theoretical visualization frameworks and their practical applicability. 
\section{User Personas}

We undertake the construction of personas' anatomical structure through a meticulous selection of constituent facets deemed conducive to the facilitation of visualization design. In our approach, we refrain from exclusive reliance on LLMs for the construction of personas. Instead, we endeavor to harness the capabilities of LLMs (synthesizing data from diverse sources, including customer feedback, surveys, interviews, and market research reports) judiciously, employing them to both construct and enrich the personas under consideration.


Analogous to other methodologies for creating personas \cite{siika2016persona,nielsen2021understanding,devitt2021creating}, this approach involves the establishment of tailored inquiries aligned with specific criteria, thereby directing attention towards predefined groups within the population. Developing data-backed personas poses significant challenges for human-computer interaction (HCI) researchers due to the substantial effort required to gather, analyze, and synthesize meaningful user data. Traditional approaches, such as surveys, interviews, and observational studies, demand considerable time, funding, and interdisciplinary expertise\cite{jung2025personacraft}. Moreover, ensuring that these personas accurately reflect diverse user experiences necessitates access to extensive datasets \cite{zhang2016data}, which are often difficult to obtain and analyze systematically. The complexity of human behavior further complicates this process, making it challenging to construct personas that are both representative and adaptable to evolving user needs. In contrast, AI-driven deep research technologies offer a more efficient and scalable alternative. By leveraging machine learning and natural language processing, these tools can process vast amounts of behavioral data, identifying meaningful patterns and user archetypes with greater speed and precision \cite{schuller2024generating}. This automation reduces the burden on researchers, allowing them to focus on interpreting insights rather than manually constructing personas, ultimately fostering more dynamic and contextually relevant design decisions.


We leverage Perplexity's advanced deep research \cite{perplexity_ai} capabilities to effectively generate personas through simple and clear prompting strategies. In this study, we focus on visualizing election data from the UK to build personas that represent various segments of the UK population. By utilizing Perplexity's deep research feature, we can process complex datasets with greater efficiency, enabling the creation of diverse user archetypes that reflect the nuances of real-world demographics. This approach allows for a comprehensive understanding of different voter segments based on their behaviors, preferences, and voting patterns, which are then visualized in an easily interpretable format. The personas generated from this data offer valuable insights into the electorate, which can help inform targeted political campaigns, social research, and policymaking. This method provides a more scalable and accurate way of segmenting and analyzing large, multifaceted datasets compared to traditional approaches like surveys or focus groups.

LLM-generated personas are good alternatives to traditional HCI researcher-developed profiles when modeling the UK electorate, primarily through enhanced efficiency and scalability. Traditional persona creation methods present prohibitive costs, logistical and ethical challenges due to privacy constraints, while often failing to capture multi-dimensional subjective attributes comprehensively \cite{li2025llm}. In contrast, LLMs provide a scalable, cost-effective alternative for generating detailed voter profiles. The resource efficiency of LLM-generated personas is particularly noteworthy, as conventional persona generation is typically not cost-efficient\cite{shin2024understanding}. While acknowledging that hybrid human-AI workflows may yield the most optimal results for certain applications, the resource advantages and scalability of LLM-generated personas make them particularly valuable for broad electoral modeling when constraints limit traditional HCI research methods.

\subsection{Anatomy of the User Personas} \label{4point1}
We define the personas by a number of characteristics,\\
\textbf{Description:}  This is a brief description of the personas upon which the anatomy is based. \\
\textbf{Technology Attitudes:} This is to discern the prevalent devices utilized by the personas and their overarching attitudes towards technology. Such investigation holds significance in comprehending the preferred medium for disseminating the visualizations. \\
\textbf{Voting Behavior:} This is to elucidate the thematic concerns that engage the interest of the personas for voting. Such investigations are pivotal for deciding the nature of the data presented within the visualizations.\\
\textbf{Information Consumption Choices:} This inquiry aims to ascertain the primary sources from which these personas acquire their news and the preferred modalities through which they consume information. Such insights are pivotal for informing design criteria pertaining to visualization.\\

\subsection{Persona Creation}




LLMs operate in accordance with the structural framework of their prompt design. This prompt design encompasses the anatomical elements of personas, thereby providing the LLM with a comprehensive template for response generation. Within the instruction, the LLM is directed to incorporate diverse demographic characteristics representative of the UK population in the creation of these personas. As previously articulated, we employ Perplexity's deep research functionality as a methodological approach to derive and develop user personas with empirical validity\cite{perplexity_deep_research}. The systematic implementation of this approach ensures that the personas reflect the heterogeneity of the target population, thereby enhancing the ecological validity of the LLM's outputs. Through this structured methodology, we facilitate the generation of responses that appropriately address the multifaceted nature of the user personas.  

The following is the prompt we use for persona creation, 
\begin{quote}
\texttt{I want you to create three user personas based on the following details,
1. They are based in the United Kingdom and are interested in election results and analysis
2. They have different attitudes towards technologies
3. The personas include; description, technology attitudes, voting behavior (what they consider while voting for a candidate/party) and information consumption choices (TV, newspaper, tablets, smartphones, etc.)
Do extensive research and present the 3 personas in the format above.}
\end{quote}

Once the personas are generated, we then use the generated persona description to generate persona images in Adobe firefly \cite{adobe_firefly}.

\begin{figure*}[!htb]
    \centering
    
    \includegraphics[width=0.95\textwidth]{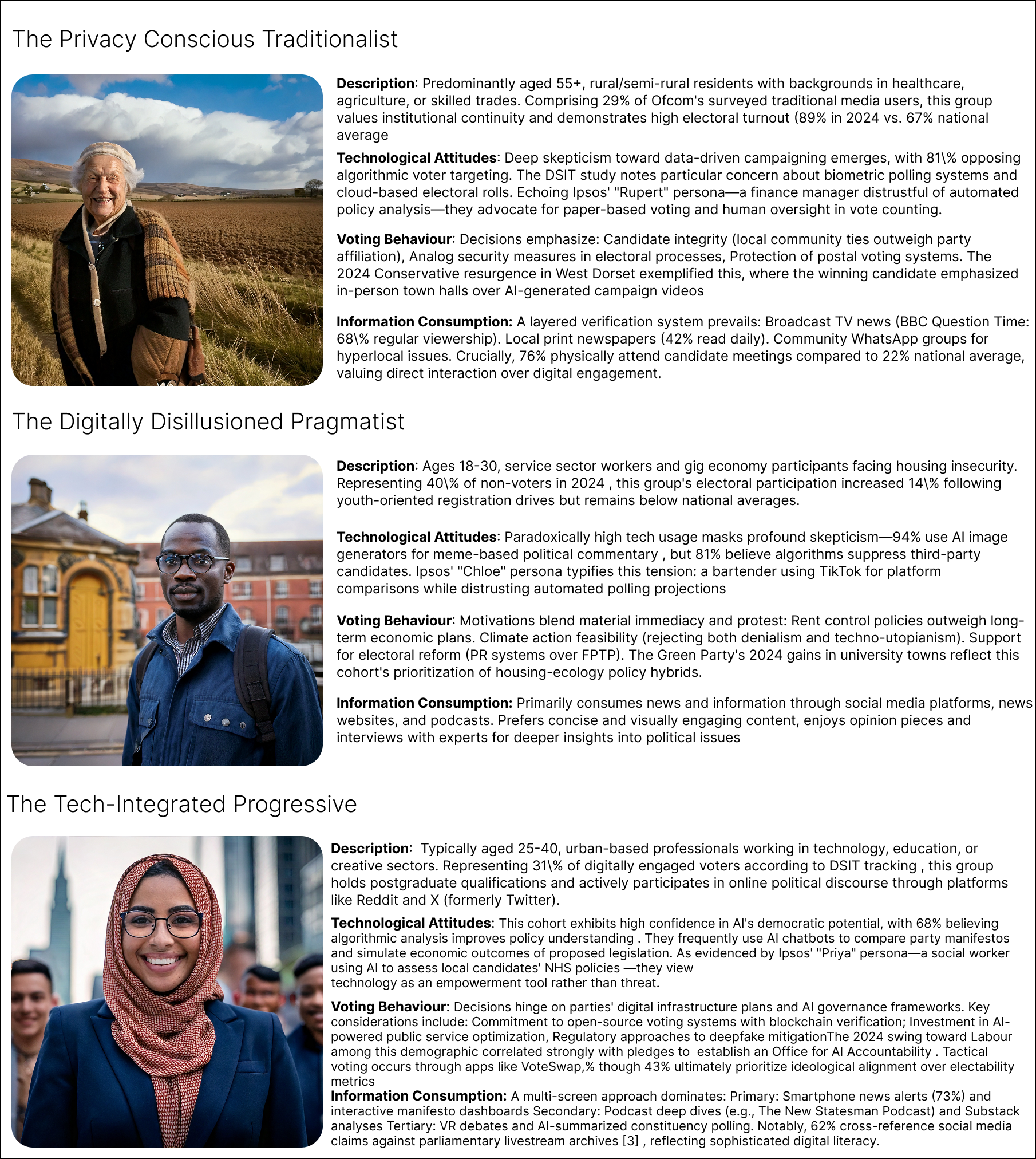}
    \caption{The 3 User Personas.}
    \label{fig:persona1}
\end{figure*}

The personas are presented in figure (\ref{fig:persona1}).



\section{Design Considerations for Visualizations}
Following the development of user personas, we established specific design considerations for visualization purposes. These considerations include selection of a visualization communication artifact, such as an infographic, video presentation, or interactive interface; chart choices and visual design such as maps, bar charts, comparative visualization or narrative visualization and data to be visualized; such as temporal data, or clustered data according to income zones. 


As described in Section (\ref{4point1}),for the design considerations, Technology Attitudes primarily guides the medium of visualization communication (like smartphones or newspapers), Information Consumption primarily guides the visualization artifact or prototype (like some visual analytics in a smartphone or infographics in a newspaper) along with the visual design and voting patterns primarily guide the choice of charts used and the type of data visualized (results on a map segmented into different economic zones, or chart of the wealth of the elected members). 

\subsection{For the privacy conscious traditionalist}
\subsubsection{Visualization Communication Artifact}\label{vca1}
Infographics effectively convey complex information in a visually engaging and easily digestible format. They simplify data and concepts, making them accessible to a broad audience, and enhance retention and engagement through combined text, graphics, and design \cite{de2018does, siricharoen2015infographic}. Visual elements illustrate relationships and trends, facilitating understanding and efficient communication, particularly in time-constrained contexts. Their aesthetic appeal promotes sharing on digital platforms as well as on print media, broadening the information's reach and impact, making infographics valuable in academic and professional fields. Static infographics serve as valuable communication tools for \textit{privacy-conscious traditionalists} by offering content through established print media channels. These visual presentations effectively deliver information while respecting privacy concerns in several key ways. Print newspapers remain trusted information sources for many traditional readers who maintain skepticism toward digital tracking mechanisms. Static infographics published through these channels cannot collect personal data, creating a natural barrier against the algorithmic targeting that concerns many voters. This inherent privacy protection helps build trust with audiences who prioritize information security and personal privacy.


\subsubsection{Chart Choices \& Visual Design}\label{ccvd1}
Effective infographic design in print integrates clarity, accuracy, and visual appeal to communicate complex information efficiently. Essential principles include: useful content with clear, concise details tailored to audience needs; legibility through readable fonts, appropriate colors, and well-labeled graphics \cite{siricharoen2015infographic,de2018does}; aesthetic quality with high-quality visuals and balanced layout; logical information structure with clear visual hierarchy; and accurate, relevant, properly sourced data.  Maps provide a spatial context that enhances users' understanding of locations, distances, and relationships between different entities, facilitating better decision-making and communication \cite{crampton2001maps}. Their visual nature leverages human cognitive abilities to recognize patterns and spatial relationships quickly, making maps more relatable and intuitive (to the \textit{privacy-conscious traditionalists}) compared to textual descriptions or data tables. Narrative visualization techniques involve the use of storytelling elements to present data in a compelling and cohesive manner, often incorporating a sequential flow that guides the viewer through the information \cite{segel2010narrative}. Comparative visualization techniques focus on superimposing or juxtaposing multiple data sets or variables to highlight differences \cite{gleicher2011visual,panda2021comparing}, similarities, and trends. This approach utilizes side-by-side charts, scatter plots, and matrix diagrams to enable direct comparisons, thereby facilitating a deeper analytical understanding of the relationships and patterns within the data. Both of these techniques leverage visual storytelling elements such as annotated charts, timelines, and flow diagrams to create a narrative arc that makes complex data more relatable and easier to understand to the \textit{privacy-conscious traditionalists}.

\subsubsection{Data}\label{d1}
\textit{Privacy-conscious traditionalists}) tend to favor larger-scale data and historical information. They value understanding trends and patterns over time (\texttt{``this group
values institutional continuity and demonstrates high electoral turnout"} - Fig \ref{fig:persona1}).  Additionally, they are often interested in outcomes at the local level, as it provides more context and relevance to their decision-making.(\texttt{``Candidate integrity (local community ties outweigh party affiliation)"} - Fig \ref{fig:persona1}) This preference reflects their focus on established data and real-world implications.



\subsection{For The Digitally Disillusioned Pragmatist}
\subsubsection{Visualization Communication Artifact}\label{vca2}
A data video is a video format that uses dynamic or animated visualizations of data to communicate information, trends, or insights \cite{yang2023understanding}. These videos typically integrate graphs, charts, infographics, and animations to transform complex datasets into digestible visual narratives, making it easier for audiences to understand and engage with the data. Data videos are particularly useful in fields like journalism, marketing, and research, where conveying large volumes of data in an accessible manner is crucial. They often combine storytelling techniques with data visualization to ensure that key insights are effectively communicated. As a powerful tool for data communication, they enable viewers to grasp patterns, correlations, and anomalies quickly, fostering informed decision-making. Data videos also have the potential to reach broader audiences due to their engaging and interactive nature, increasing data literacy and promoting better comprehension of statistical information. Data videos represent an effective communication format for platforms like TikTok in mobile devices where \textit{younger audiences} predominantly gather. These short-form video presentations offer several strategic advantages for information dissemination in today's digital landscape. The concise nature of data videos aligns perfectly with the viewing preferences on social media platforms where attention spans are limited. By condensing complex information into brief, visually compelling presentations, these videos can effectively capture viewer interest and convey key messages efficiently.


\subsubsection{Chart Choices \& Visual Design} \label{ccvd2}
Data visualization for mobile devices necessitates a design that emphasizes simplicity, clarity, and responsiveness due to the limited screen real estate and varying user contexts. Key characteristics include using minimalistic and intuitive graphical elements to avoid overwhelming the user, ensuring that visualizations are easily interpretable at a glance \cite{lee2018data, meyer2016visualization,coleti2019design}. These visualizations should be responsive, adapting seamlessly to different screen sizes and orientations, and should maintain performance efficiency to accommodate varying device capabilities and network conditions. Animated charts enhance data videos through their dynamic visualization capabilities, with bar chart races serving as a prime example. A bar chart race \cite{flourish_bar_chart_race} displays the changing values of different categories over time, allowing viewers to observe progression, regression, and shifting rankings in a compelling visual format. The animation creates a narrative arc as bars extend or contract, capturing attention through movement and competitive positioning. Maps likewise excel in data visualization due to their inherent familiarity and geographic context, providing audiences with spatial reference points that connect abstract data to recognizable locations. Both visualization types leverage animation and relatability to transform complex datasets into accessible, engaging stories that resonate with viewers and improve information retention.

\subsubsection{Data}\label{d2}
\textit{Younger audiences }are increasingly focused on data related to national-scale, ideological, climate action, green energy, and similar topics. They prefer information that is presented in a clear, concise manner. Short videos are particularly appealing to them, as they simplify complex concepts \texttt{``Viral TikTok explainers (average 12-second engagement per clip)" - Fig (\ref{fig:persona1})}. This format allows them to quickly grasp important points without feeling overwhelmed. Overall, they value easily digestible data that doesn't require in-depth analysis.
\begin{figure*}[!htb]
    \centering
    \fbox{
    \includegraphics[width=0.92\textwidth]{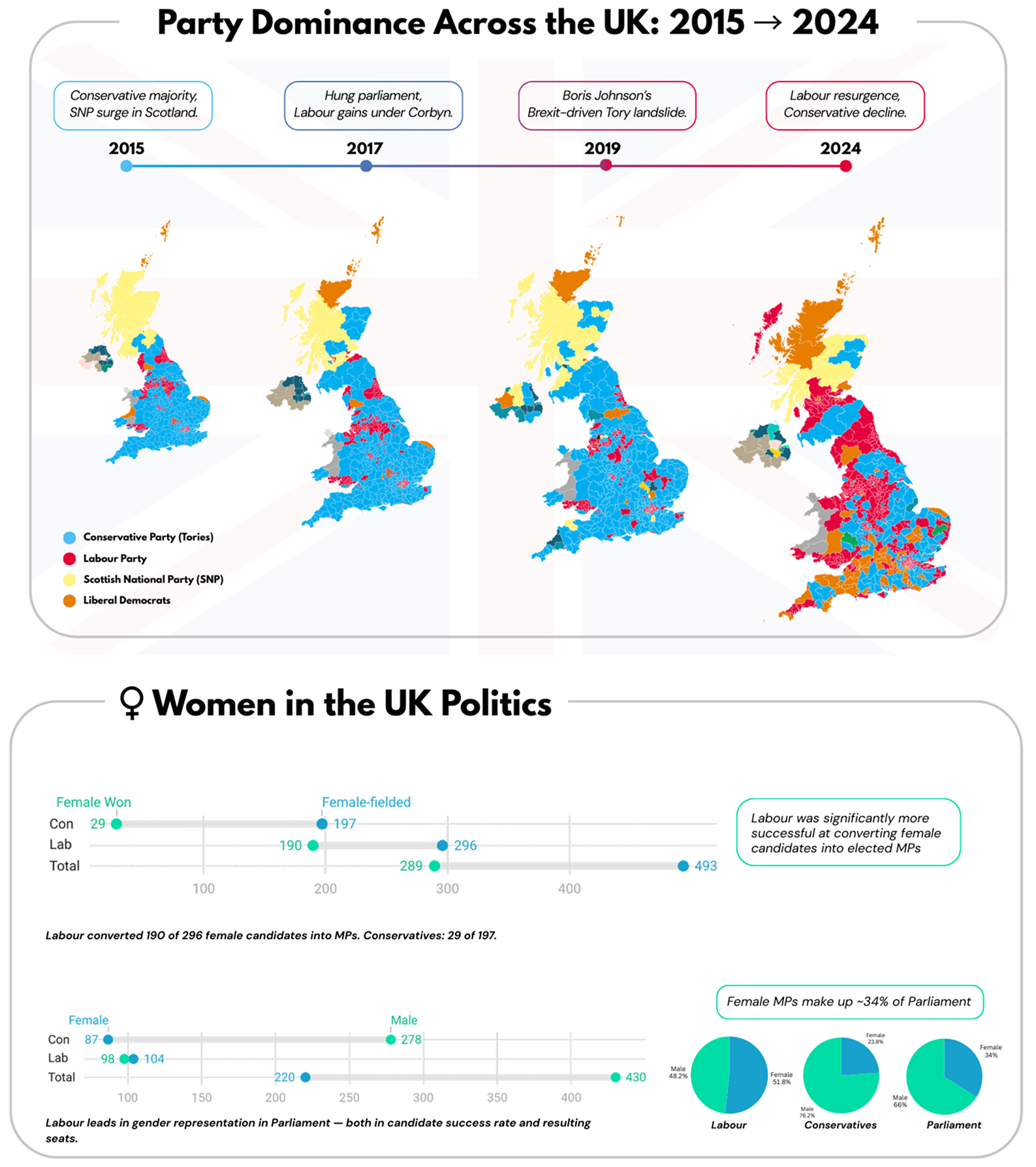}}
    \caption{A infographic prototype.}
    \label{fig:infographic}
\end{figure*}

\begin{figure*}[!h]
  \centering
  \begin{minipage}[b]{0.32\textwidth}
    \includegraphics[width=\textwidth]{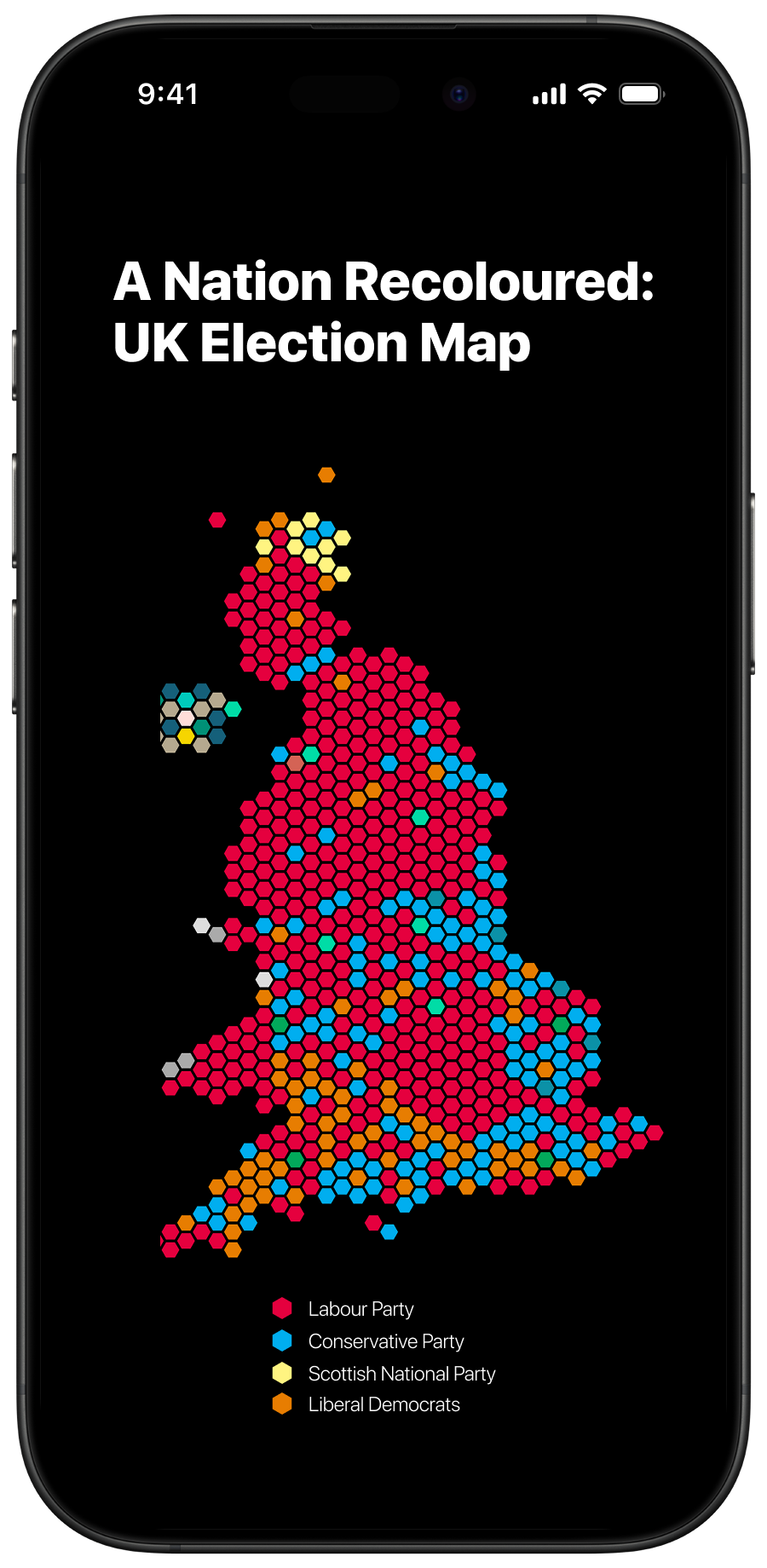}
  \end{minipage}
  \hfill
  \begin{minipage}[b]{0.32\textwidth}
    \includegraphics[width=\textwidth]{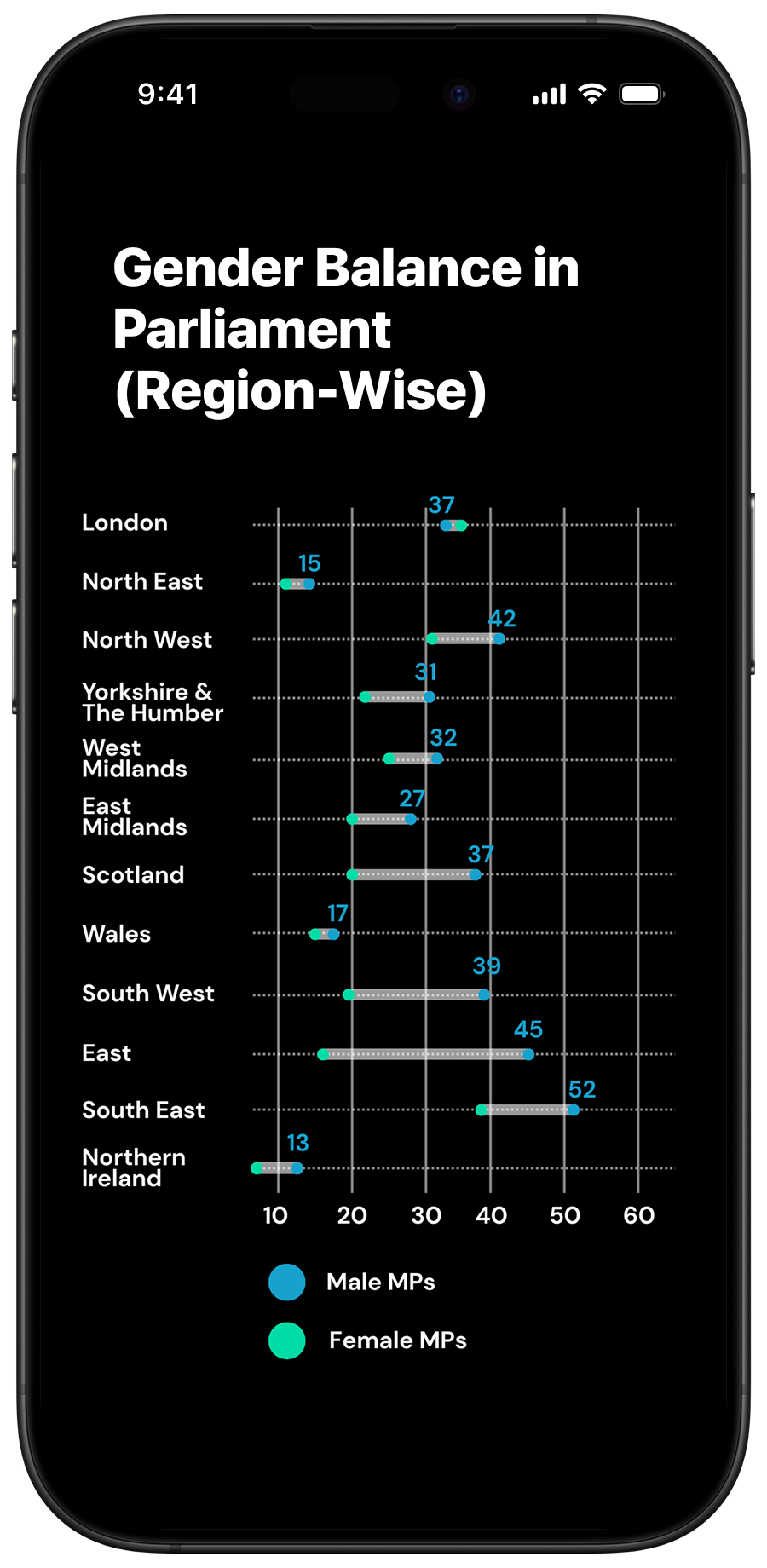}
  \end{minipage}
    \hfill
  \begin{minipage}[b]{0.32\textwidth}
    \includegraphics[width=\textwidth]{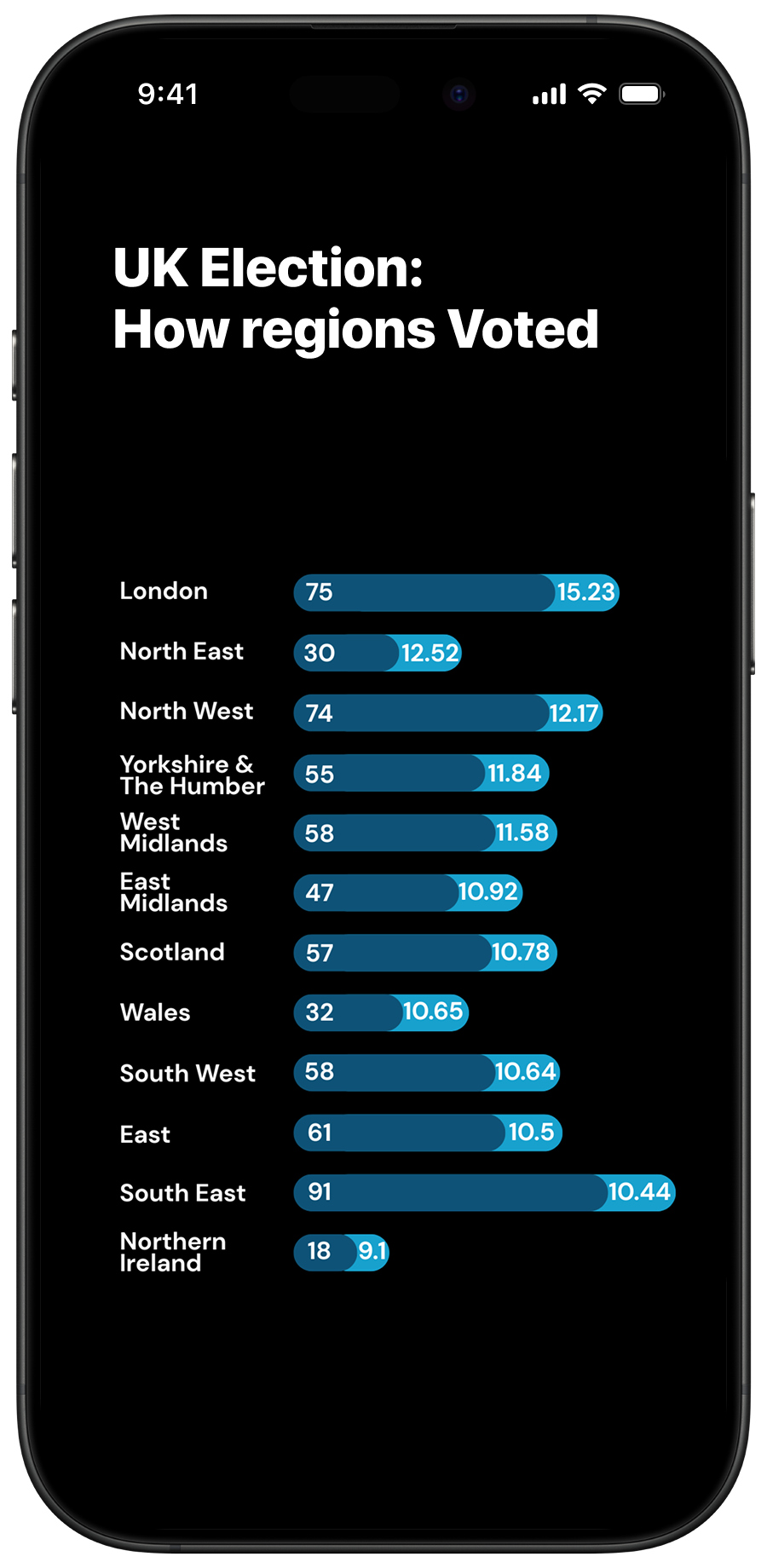}
  \end{minipage}
  \caption{Smartphone prototypes}
  \label{fig:smartphone}
\end{figure*}

\begin{figure*}[!tbp]
  \centering
  \begin{minipage}[b]{0.24\textwidth}
    \includegraphics[width=\textwidth]{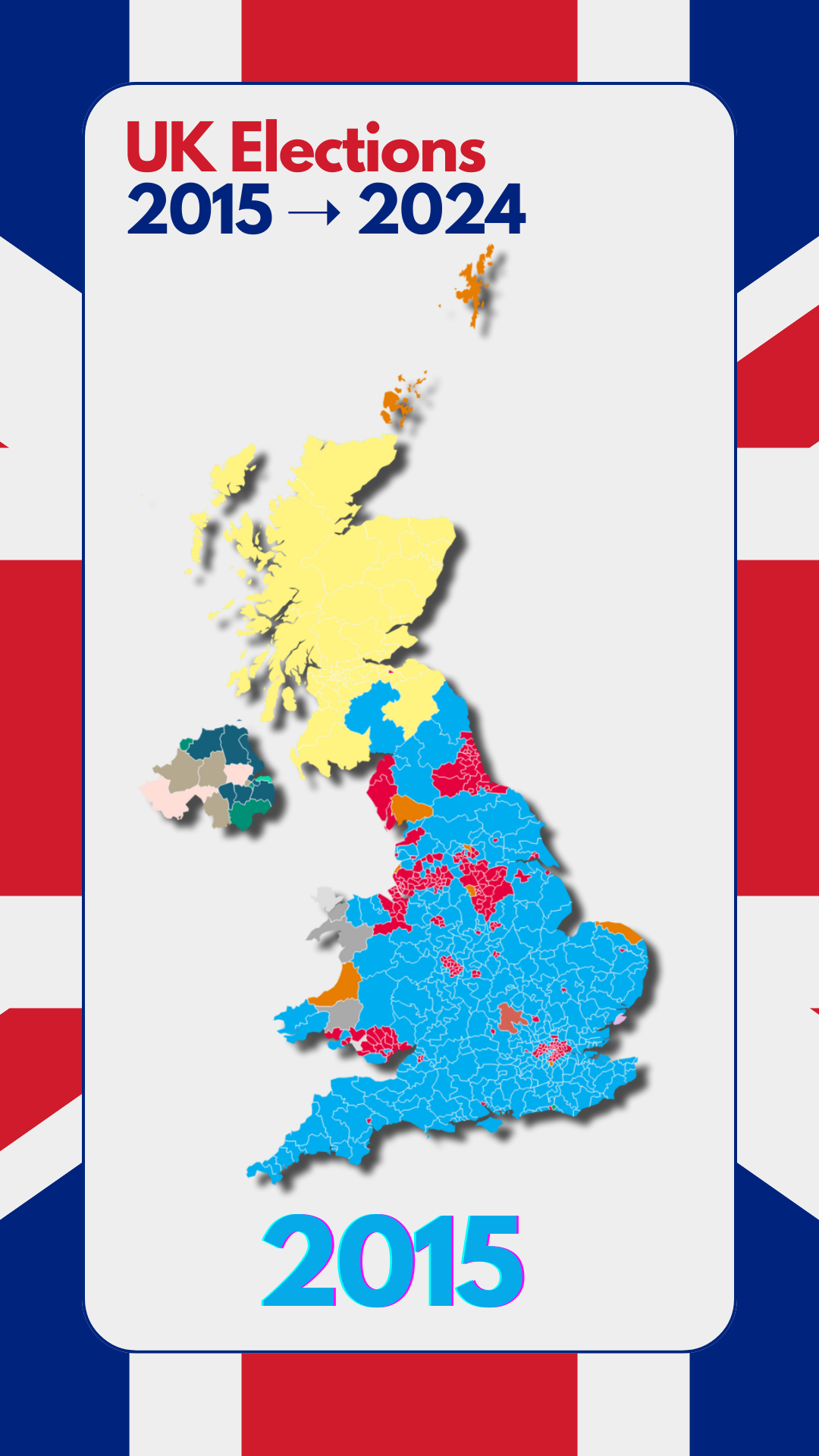}
  \end{minipage}
  \hfill
  \begin{minipage}[b]{0.24\textwidth}
    \includegraphics[width=\textwidth]{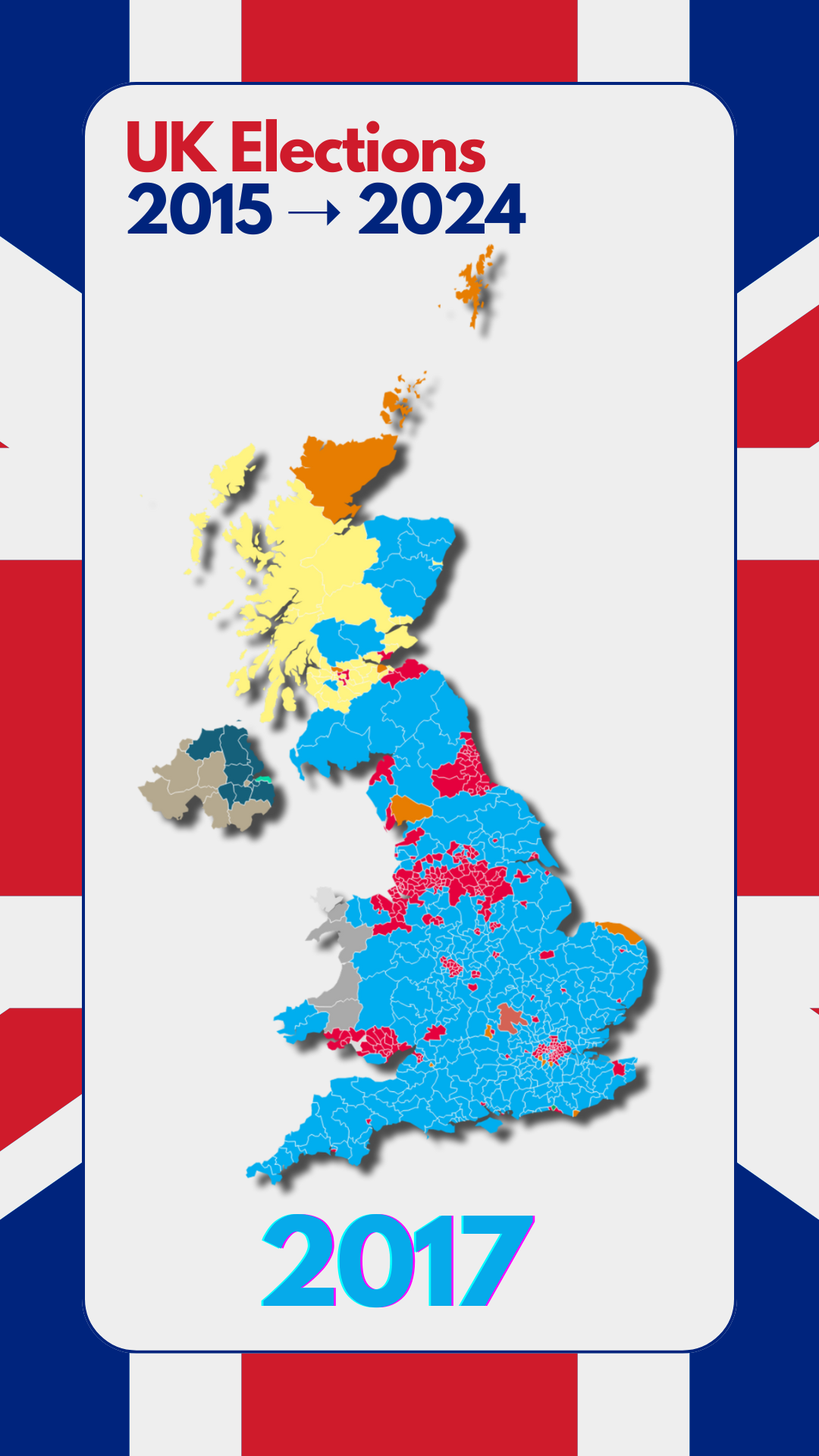}
  \end{minipage}
    \hfill
  \begin{minipage}[b]{0.24\textwidth}
    \includegraphics[width=\textwidth]{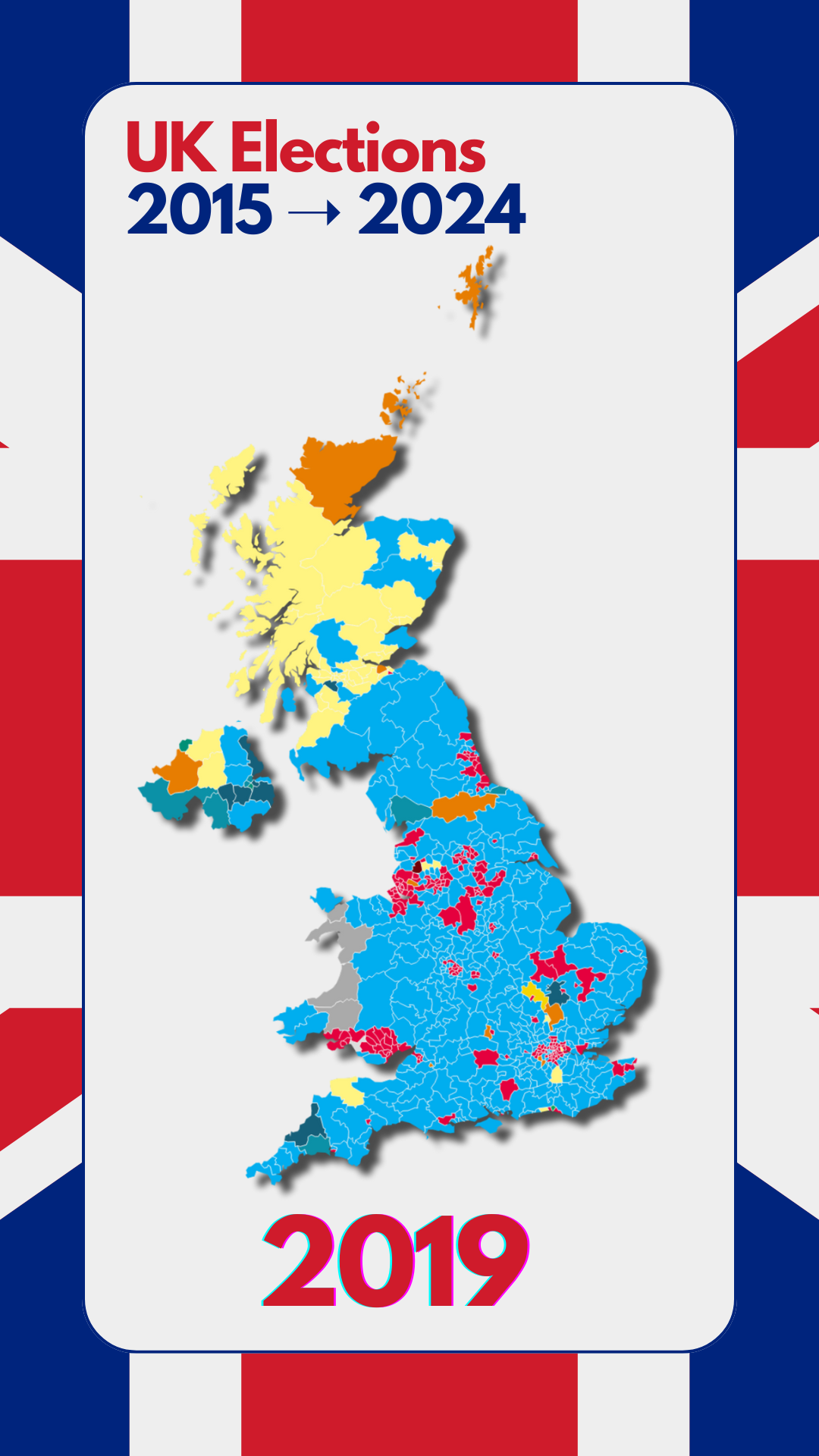}
  \end{minipage}
    \hfill
  \begin{minipage}[b]{0.24\textwidth}
    \includegraphics[width=\textwidth]{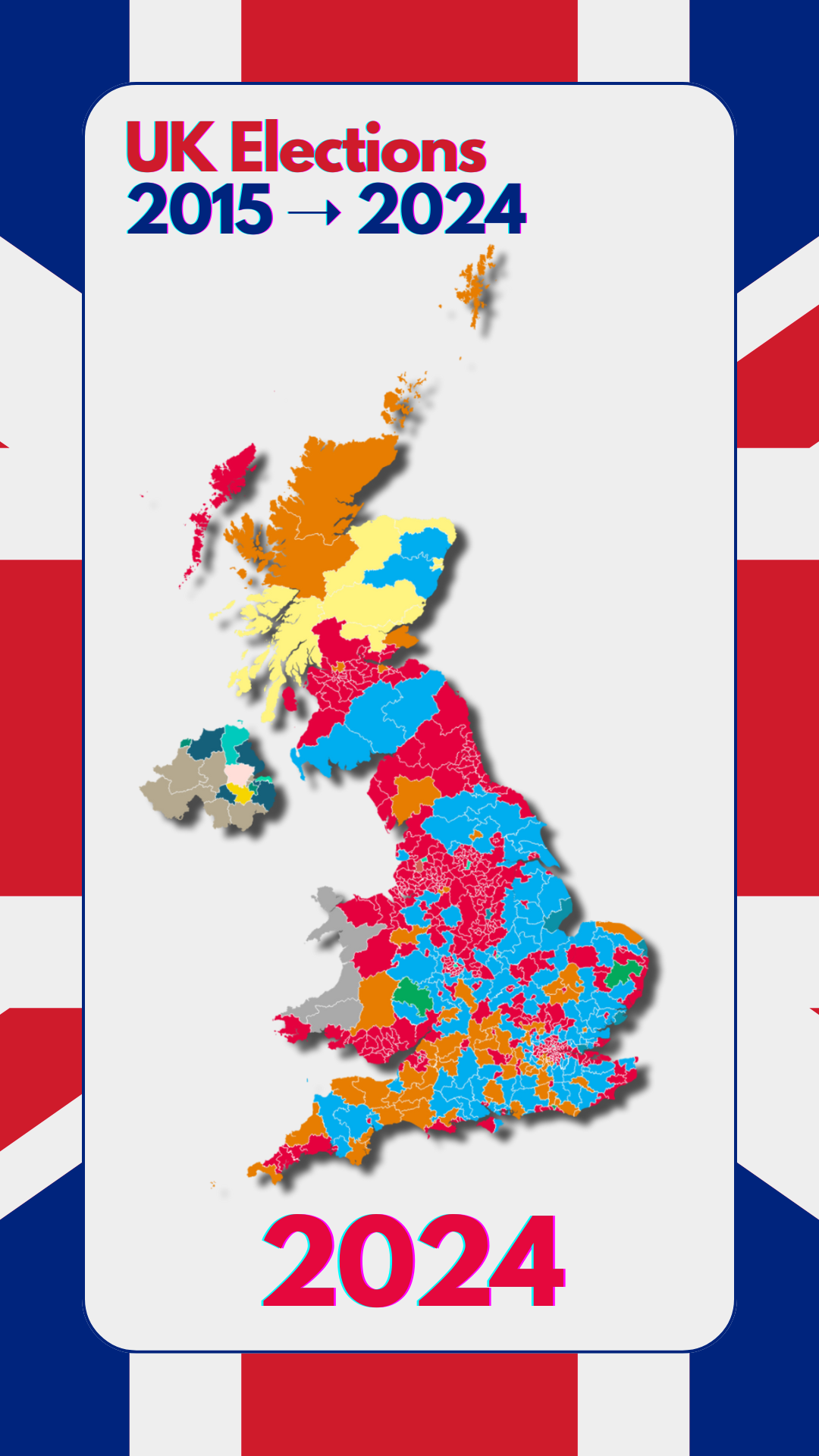}
  \end{minipage}
  \caption{Four snapshots of the Data Video prototype}
  \label{fig:datavideo}
\end{figure*}

\subsection{For the Tech Integrated Progressive}
\subsubsection{Visualization Communication Artifact}\label{vca3}
Interactive visualization serves as a powerful tool in academic research and data analysis, offering significant utility by enhancing user engagement and comprehension of complex datasets. By allowing users to manipulate and explore data dynamically, interactive visualizations facilitate deeper insights, foster greater understanding, and uncover hidden patterns that static representations might obscure \cite{keim2008visual}. They enable the examination of multiple dimensions and variables simultaneously, providing a more comprehensive perspective to the \textit{``Tech Integrated Progressive"}. Additionally, interactive visualization promotes accessibility and user-driven inquiry, empowering researchers and stakeholders to test hypotheses, identify correlations, and make data-driven decisions with greater confidence. This adaptability not only augments the analytical process \cite{wong2004visual} but also supports more effective communication of findings, making complex data more accessible and interpretable to diverse audiences.

\subsubsection{Chart Choices \& Visual Design}\label{ccvd3}
When creating visualizations for digital devices, it is crucial to prioritize clarity and readability, ensuring that the design is both aesthetically pleasing and functional. It is important to utilize a responsive design approach to accommodate various screen sizes and resolutions, ensuring that the visual content remains legible and well-organized across different devices \cite{siricharoen2015infographic}. A consistent color scheme with adequate contrast helps improve visibility and draw attention to key data points, while avoiding too many colors that can cause strain. Simplifying visual elements allows users to focus on the most relevant information, and interactive features should be used sparingly to keep engagement without overwhelming them. Advanced data charts, such as heat maps, radar charts, and network diagrams, are invaluable tools for comprehensively analyzing complex datasets. These sophisticated visualizations facilitate the identification of patterns, trends, and correlations that might remain obscured in traditional charts. Heat maps, for instance, use color gradients to represent data density and variations, making them particularly effective for large datasets. Radar charts, or spider charts, enable the comparison of multiple variables across different entities, offering a multi-dimensional perspective \cite{wong2004visual}. Network diagrams, essential in fields like sociology and computer science, visually represent relationships and flows between interconnected nodes. The use of these advanced charts enhances data interpretation by providing intuitive and insightful representations, thereby supporting more informed decision-making and fostering a deeper understanding of intricate data structures for the \textit{``Tech-Integrated Progressive"}

\subsubsection{Data} \label{d3}

Tech-integrated progressives seek to analyze data at a deeper level, focusing on detailed insights while also maintaining a broader perspective. They aim to merge in-depth data analysis with an understanding of the overall picture. This approach allows for a more comprehensive view that considers both the specifics and the larger trends. By integrating technology, they enhance their ability to gather, interpret, and apply data effectively. Ultimately, this balance helps them make more informed decisions(\texttt{``Investment in AI-powered public service optimization - Regulatory approaches to deepfake mitigation"} - Fig \ref{fig:persona1}) in complex situations.

\section{Visualization Prototypes from Design Principles}

In order to showcase the practical utility and effectiveness of the design considerations, we undertook the development of visualization prototypes derived from these criteria. These prototypes serve as representations, providing a visual and interactive means to evaluate how the theoretical design considerations translate into functional and user-centered (as the considerations were derived from user personas) visualizations. By anchoring the design principles in practical prototypes we create a foundation for further exploration and validation of their relevance in diverse real-world scenarios. 

We refer to the design considerations used to develop the prototypes in-text.


\subsection{For the privacy conscious traditionalist}
This prototype (Fig \ref{fig:infographic}) presents a chronological visualization of UK parliamentary elections from 2015 to 2024. By employing choropleth maps (\ref{ccvd1}), the electoral performance of various political parties (\ref{d1}) is systematically compared across successive election cycles \cite{gleicher2011visual}. Additionally, the infographic (\ref{vca1}) illustrates the electoral representation of women in UK politics through range plots, depicting the number of women who contested elections versus those who secured seats (\ref{d1}). Another range plot examines the gender composition of the UK Parliament by comparing the number of male and female MPs (\ref{d1}). Furthermore, pie charts have been incorporated to depict proportional distributions. This infographic is designed for potential publication in newspapers (\ref{vca1}), accompanied by a dataset and an analytical report. The charts for this prototype were generated using Datawrapper, while the infographic was designed in Canva. The data was sourced from the UK government website \cite{UKParliament2023}.


 \subsection{For the Digitally Disillusioned Pragmatist}
This prototype (Fig \ref{fig:datavideo}) presents a chronological visualization of UK parliamentary election results (\ref{d2}) spanning the years 2015 to 2024. The color schemes are consistently applied, with individual colors (\ref{ccvd2}) derived from the branding colors of political parties. The maps are uniformly sized (\ref{ccvd2}) to facilitate effortless comparison of results. This prototype takes the form of a data video (\ref{vca2}), designed for dissemination on social media platforms such as TikTok (\ref{vca2}). The video is brief and requires minimal user attention (\ref{ccvd2}). The maps were initially created in Datawrapper, and the final video was designed using Canva. The data was sourced from the UK government website \cite{UKParliament2023}.


 \subsection{For the Tech Integrated Progressive}

This prototype (Fig \ref{fig:smartphone}) is an interactive visualization (\ref{ccvd3}) that provides details on demand (\ref{ccvd3}). It focuses on the data from the UK 2024 elections, offering an in-depth analysis of how different regions of the UK voted (\ref{d3}). Additionally, it includes a section on gender balance in the UK Parliament, categorized by region (\ref{d3}). The visualization employs hexagonal bin charts to represent election results, while bar charts and range plots are used to provide further insights into the data (\ref{ccvd3}). The charts were created using Datawrapper, and the prototypes were designed using Figma.

\section{Prototype Evaluation}

We combine LLMs with heuristic evaluation methods to assess visualization prototypes more comprehensively. This dual approach leverages LLMs' natural language processing capabilities to provide nuanced feedback that complements traditional heuristic evaluations \cite{siricharoen2015infographic} based on established principles and expert guidelines. We believe LLMs can enhance heuristic evaluation by offering additional insights not immediately evident through conventional analysis. This integrated methodology aims to identify usability issues in the design process (eg low fidelity prototypes), ultimately improving the quality and user experience of visualizations.

The three prototypes are provided to the LLM in the form of images (or sequence of images) to maintain consistency across all the prototypes. To verify the accurate interpretation of the image, the LLM is prompted to generate a descriptive summary of its contents. Subsequently, the LLM is queried to assess whether the visualization aligns with established visualization heuristics \cite{siricharoen2015infographic,kuparinen2013introducing} and the user persona description (\ref{fig:persona1}). To ensure privacy, the evaluation is conducted using the temporary ChatGPT mode. 



\subsection{Infographic}

For the infographic, the LLM's summarized response included,
\begin{quote}
    \centering
    \texttt{The infographic effectively conveys UK political trends (2015-2024) and gender representation using clear visuals, accurate data, and intuitive design. It excels in readability, organization, and audience appropriateness, employing distinct colors and labels. The structured layout ensures easy navigation, making it an informative and engaging representation of electoral and demographic changes.}
\end{quote}

\subsection{Data Video}
For the Data Video, the LLM's summarized response included,
\begin{quote}
    \centering
    \texttt{The visualization effectively maintains consistency in layout and design, aiding interpretability. However, potential issues include the changing distribution of colors, which may require a clear legend for precise interpretation. The background pattern, though patriotic, may introduce visual clutter. Additionally, uniform sizing of maps ensures comparability, but minor distortions in boundaries could impact perception.}
\end{quote}

\subsection{Interactive Visualization}
For the interactive visualization, the LLM's summarized response included,
\begin{quote}
    \centering
    \texttt{The prototype aligns well with Kuparinen et al.'s usability heuristics for mobile map applications, excelling in visibility, consistency, and minimalist design. However, it could improve user control, error prevention, and flexibility by incorporating interactive feedback, navigation controls, and tooltips. Adding documentation and interactive elements would enhance usability, ensuring better engagement and comprehension of election data.}
\end{quote}

\section{Discussion}

Our work draws heavily from visualization practice, particularly the design considerations and development of prototypes. We present a model (Fig. \ref{fig:framework}) \cite{munzner2008process} and then demonstrate the working of the model. Our work shares notable similarities with customized and personalized visualization approaches \cite{mutlu2016vizrec,qian2022personalized}, which focus on tailoring visual representations to the specific needs and preferences of individual users. Much like personalized visualizations, our approach emphasizes adapting the design elements based on contextual factors, such as the user’s goals, the nature of the data, and the desired insights\cite{pruitt2003personas}. By considering these factors, we aim to create more relevant and user-centric visualizations. 


Elections hold significant importance in democratic societies, serving as a fundamental mechanism through which citizens express their preferences and influence governance. The interpretation of election data, however, varies considerably among different audiences, influenced by diverse factors such as political affiliations, socio-economic backgrounds, and informational needs \cite{angelucci2024journalistic}. Consequently, the field of visualization research has a critical role to play in effectively catering to these varied audiences\cite{correll2019ethical}. By developing sophisticated and nuanced visual representations of election data, researchers can enhance the accessibility, comprehensibility, and engagement of electoral information for a broad spectrum of users. 
From the study we derived some actionable insights for visualization researchers. 
\subsection{Actionable Insights}
\begin{enumerate}
    \item User personas play a critical role in personalizing data visualizations by tailoring outputs to individual needs and preferences. By defining archetypal users, these personas enable the design of visualization systems that account for varying levels of expertise, goals, and contexts. Integrating personas with visualization tools enhances user engagement, as the resulting displays align more closely with specific interpretive requirements. However, the effectiveness of this approach depends on accurately capturing user diversity and translating it into adaptable visual frameworks. As visualization technologies evolve, leveraging user personas promises to refine personalization, ensuring that data representations are both accessible and relevant across diverse user groups.
    \item LLMs enable a intermediate steps (such as creating personas, or evaluating visualization guided by heuristics) in data visualization, bridging raw datasets and interpretable outputs. By processing queries and identifying patterns, LLMs generate preliminary visuals, enhancing accessibility to complex information. Though this marks progress, advanced LLM capabilities—such as autonomously creating sophisticated, context-aware visualizations—remain under development. Current models facilitate analysis but lack the autonomy and refinement anticipated in future iterations. As research advances, these enhanced features promise to streamline data interpretation and deepen insights across academic and professional fields, highlighting LLMs’ evolving role in transforming how datasets are visualized and understood.
    \item LLMs, in their current form, are not replacing the roles of designers in visualization. Instead, they are occupying a distinct position within the visualization ecosystem. Rather than asking LLMs to evaluate visualizations using their own we deliberately instruct them to assess visualizations according to established heuristics and principles. This strategic approach creates a new visualization paradigm where LLMs serve as specialized assistants with defined responsibilities. As a consequence, the traditional role of designers will transform—shifting toward curation, direction, and refinement rather than execution of every detail. Similarly, users may experience changes in their engagement with visualizations, potentially becoming more involved in the conceptual and evaluative processes while relying on LLMs for implementation. The introduction of LLMs as independent agents in visualization workflows represents a significant evolution in how we conceive, create, and interact with visualizations.

\end{enumerate}
\subsection{Limitations \& Future Work}

This study presents a visualization framework that embraces user-centered principles without direct user participation. We argue that merely involving selected users does not inherently constitute a user-centered design process \cite{linxen2021weird,jena2021next}. Our core contribution lies in incorporating user perspectives through carefully developed user personas. Future work includes the exploration of more detailed methodologies for persona generation by LLMs, which may encompass the refinement of prompting strategies or the implementation of thematic analysis techniques on different data sources \cite{oleaevaluating}. Such advancements aim to enhance the precision and granularity of persona creation processes, thereby augmenting their utility in informing user-centered visualization design. More detailed design criteria and considerations can also be derived from such advanced personas. Further we aim to develop increasingly intricate and refined visualization prototypes across different mediums of dissemination, aligning with the design considerations extrapolated from these personas.

This study utilizes low-fidelity prototypes to demonstrate the practical application of the proposed framework. This was evident in the LLM-based evaluation as well, \texttt{``minor
distortions in boundaries", ``it could improve user control, error
prevention, and flexibility "}. Future research will include improving these prototypes and focus on developing high(er)-fidelity visualization prototypes with enhanced detail and sophistication. We can then apply established visualization heuristics and criteria from various dissemination mediums, such as newspapers, TikTok videos, or mobile apps, to assess these prototypes using LLMs.


\section{Conclusion}


In this work, we proposed the utilization of personas as a methodological approach for visualizing election data. Leveraging LLMs, we constructed three personas representative of the diverse user segments within the United Kingdom. These personas formed the basis for establishing design considerations. We then used these considerations to develop visualization prototypes, for dissemination in different mediums like infographics, data videos and interactive visualizations. We evaluated these prototypes using LLMs and presented some actionable insights for visualization researchers. 

\bibliographystyle{tfq}
\bibliography{interacttfqsample}



\section*{Supplementary material}

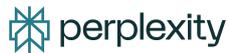

# AI-Driven Voter Personas: Technological Attitudes and Electoral Engagement in the United Kingdom

Recent research into voter behavior reveals a complex interplay between technological adoption, information consumption patterns, and electoral decision-making. Analysis of Ipsos AI personas[1], DSIT's public attitude surveys[2], and Ofcom's media plurality study[3] demonstrates that voters increasingly navigate political landscapes through distinct digital and analog lenses. This report identifies three archetypal voter profiles emerging in the UK's 2024 political environment, differentiated by their relationship with technology, prioritized policy issues, and media engagement strategies. These personas—the Tech-Integrated Progressive, the Privacy-Conscious Traditionalist, and the Digitally Disillusioned Pragmatist—reflect divergent responses to AI-driven campaigning, data privacy concerns, and hybrid information ecosystems shaping modern democracy.

## The Tech-Integrated Progressive

### Demographic Profile

Typically aged 25-40, urban-based professionals working in technology, education, or creative sectors. Representing 31% of digitally engaged voters according to DSIT tracking[2:1], this group holds postgraduate qualifications and actively participates in online political discourse through platforms like Reddit and X (formerly Twitter).

### Technological Attitudes

This cohort exhibits high confidence in AI's democratic potential, with 68% believing algorithmic analysis improves policy understanding[2:2]. They frequently use AI chatbots to compare party manifestos and simulate economic outcomes of proposed legislation. As evidenced by Ipsos' "Priya" persona—a social worker using AI to assess local candidates' NHS policies[1:1]—they view technology as an empowerment tool rather than threat.

### Voting Behavior

Decisions hinge on parties' digital infrastructure plans and AI governance frameworks. Key considerations include:

- Commitment to open-source voting systems with blockchain verification
- Investment in AI-powered public service optimization
- Regulatory approaches to deepfake mitigation

The 2024 swing toward Labour among this demographic correlated strongly with pledges to establish an Office for AI Accountability [2:3]. Tactical voting occurs through apps like VoteSwap, though 43% ultimately prioritize ideological alignment over electability metrics [1:2].

## Information Consumption

A multi-screen approach dominates:

- Primary: Smartphone news alerts (73%) and interactive manifesto dashboards
- Secondary: Podcast deep dives (e.g., *The New Statesman Podcast*) and Substack analyses
- Tertiary: VR debates and AI-summarized constituency polling

Notably, 62% cross-reference social media claims against parliamentary livestream archives [3:1], reflecting sophisticated digital literacy.

## The Privacy-Conscious Traditionalist

### Demographic Profile

Predominantly aged 55+, rural/semi-rural residents with backgrounds in healthcare, agriculture, or skilled trades. Comprising 29% of Ofcom's surveyed traditional media users [3:2], this group values institutional continuity and demonstrates high electoral turnout (89% in 2024 vs. 67% national average [1:3]).

### Technological Attitudes

Deep skepticism toward data-driven campaigning emerges, with 81% opposing algorithmic voter targeting [2:4]. The DSIT study notes particular concern about biometric polling systems and cloud-based electoral rolls. Echoing Ipsos' "Rupert" persona—a finance manager distrustful of automated policy analysis [1:4]—they advocate for paper-based voting and human oversight in vote counting.

### Voting Behavior

Decisions emphasize:

- Candidate integrity (local community ties outweigh party affiliation)
- Analog security measures in electoral processes
- Protection of postal voting systems

The 2024 Conservative resurgence in West Dorset exemplified this, where the winning candidate emphasized in-person town halls over AI-generated campaign videos [1:5].

## Information Consumption

A layered verification system prevails:

1. Broadcast TV news (BBC Question Time: 68% regular viewership [3:3])
2. Local print newspapers (42% read daily [3:4])
3. Community WhatsApp groups for hyperlocal issues

Crucially, 76% physically attend candidate meetings compared to 22% national average [1:6], valuing direct interaction over digital engagement.

## The Digitally Disillusioned Pragmatist

### Demographic Profile

Ages 18-30, service sector workers and gig economy participants facing housing insecurity. Representing 40% of non-voters in 2024 [1:7], this group's electoral participation increased 14% following youth-oriented registration drives but remains below national averages.

### Technological Attitudes

Paradoxically high tech usage masks profound skepticism—94% use AI image generators for meme-based political commentary [2:5], but 81% believe algorithms suppress third-party candidates [1:8]. Ipsos' "Chloe" persona typifies this tension: a bartender using TikTok for platform comparisons while distrusting automated polling projections [1:9].

### Voting Behavior

Motivations blend material immediacy and protest:

- Rent control policies outweigh long-term economic plans
- Climate action feasibility (rejecting both denialism and techno-utopianism)
- Support for electoral reform (PR systems over FPTP)

The Green Party's 2024 gains in university towns reflect this cohort's prioritization of housing-ecology policy hybrids.

### Information Consumption

A fragmented media diet emerges:

- Viral TikTok explainers (average 12-second engagement per clip)
- Public library access to bypass paywalls (32% utilization [3:5])
- Street art/murals as political messaging vectors

Notably, 68% use ad blockers to avoid targeted political ads [2:6], while 55% participate in "info-swaps" through encrypted messengers like Signal.

## Technological Determinants of Voting Patterns

### AI's Role in Candidate Evaluation

The Tech-Integrated cohort utilizes natural language processing tools to compare manifesto commitments across election cycles. For instance, 58% employed sentiment analysis APIs to detect inconsistencies in candidates' climate statements[2:7]. Conversely, Traditionalists dismiss such tools as "artificial authenticity," favoring handwritten candidate responses to constituent letters[1:10].

### Data Privacy Tradeoffs

DSIT's tracking reveals a stark divide: 73% of Progressives accept personalized campaign analytics in exchange for predictive policy simulators, while 89% of Traditionalists demand opt-out defaults for voter profiling[2:8]. Pragmatists exhibit bifurcated behavior—sharing TikTok viewing habits freely but using burner emails for voter registration[1:11] [3:6].

### Misinformation Navigation Strategies

Progressives rely on blockchain-verified fact-checking consortiums, Traditionalists on BBC Verify annotations, and Pragmatists on crowd-sourced credibility metrics through Reddit's r/UKPolitics community[3:7]. Notably, all groups now bypass Google Search for political queries due to perceived filter bubble manipulation[3:8].

## Media Ecology and Electoral Influence

### Second Screening Dynamics

Ofcom data shows 62% of Progressives use tablets to fact-check TV debates in real-time, compared to 9% of Traditionalists[3:9]. This parallel information processing creates divergent debate perceptions—Progressive viewers emphasized climate policy timestamps, while Traditionalists focused on candidates' demeanor under questioning[1:12].

### Local vs National Source Credibility

Traditionalists grant local newspapers 78% trust ratings vs 22% for national outlets[3:10]. Progressives exhibit inverse trust patterns (64% national/36% local), while Pragmatists equally distrust all institutional media (82%[3:11]).

### Analog Resurgence

Contrary to tech-dominant narratives, 2024 saw 41% increase in printed manifesto requests across age groups[1:13], suggesting hybrid digital-physical strategies. The Labour campaign's QR-coded lawn signs—linking to both AI summaries and printable PDFs—exemplified this synthesis[1:14] [2:9].

**Conclusion: Technological Polarization and Democratic Resilience**

The 2024 UK electoral landscape reveals not technological homogenization, but rather the crystallization of distinct digital-physical voter identities. While Progressives drive adoption of predictive polling interfaces and decentralized voting prototypes, Traditionalists anchor electoral integrity in human-mediated processes. Pragmatists navigate this spectrum through ironic detachment and subversive tech appropriation.

Emerging challenges include:

- Mitigating algorithmic exclusion of analog-preference voters
- Developing hybrid campaign formats satisfying screen-native and paper-native constituents
- Preventing privacy paradoxes where high-tech users become de facto electoral arbiters

Future research must assess whether these personas represent stable political identities or transitional adaptations to accelerating technological change. The DSIT survey's finding that 63% of Pragmatists would support AI-designed "policy cocktails" blending rival manifestos[2:10] suggests potential paradigm shifts in issue-based voting.

Ultimately, UK democracy's resilience may hinge on creating interoperable systems accommodating all three technological engagement modes—a challenge requiring equal parts innovation and institutional restraint.

<div align="center">❄</div>



\end{document}